\documentclass[reprint,amsmath,amssymb,aps,pre,floatfix]{revtex4-2}

\usepackage{graphicx}
\usepackage{dcolumn}
\usepackage{bm}
\usepackage{hyperref}
\usepackage{xcolor}

\begin{document}

\title{Long-range social pressure and the evolution of cooperation in multiplex networks}

\author{Mar\'ia Pereda}
\email{maria.pereda@upm.es}
\affiliation{Grupo de Investigaci\'on Ingenier\'ia de Organizaci\'on y Log\'istica (IOL),
  Departamento de Ingenier\'ia de Organizaci\'on, Administraci\'on de Empresas y Estad\'istica,
  Escuela T\'ecnica Superior de Ingenieros Industriales,
  Universidad Polit\'ecnica de Madrid, 28006 Madrid, Spain}
\affiliation{Grupo Interdisciplinar de Sistemas Complejos (GISC), 28911 Legan\'es, Madrid, Spain}

\author{Gabrielle Muller}
\affiliation{Universidad Polit\'ecnica de Madrid, 28006 Madrid, Spain}

\date{\today}

\begin{abstract}
Social pressure---the awareness of being observed by others---is a fundamental driver of
prosocial behavior in human societies.
Yet it is typically assumed that only direct neighbors exert vigilance pressure on an
individual, despite empirical evidence that social influence persists to at least three
degrees of separation.
Here we show that extending the reach of social vigilance beyond direct neighbors
substantially promotes cooperation.
We couple a Prisoner's Dilemma on one layer of a multiplex network to a vigilance cascade
on the other, with influence decaying geometrically with network distance.
Extending vigilance to just the second circle of influence ($L=2$) shifts the critical
temptation for defection by roughly 36\% in sparse Erd\H{o}s--R\'enyi networks.
Extending to four circles ($L=4$) raises this threshold by over 50\% in the same
networks.
Both shifts are topology dependent: the $L=1\to2$ step alone already accounts for most
of the gain in these sparse networks, whereas the pattern is more gradual in denser,
hub-dominated Barab\'asi--Albert networks, consistent with the decay coefficients of
social influence reported in controlled experiments.
The effect is strongest in sparse networks, requires that the vigilance and game layers
be aligned, and reproduces directly on a real social network of physicians; in dense,
hub-dominated networks the gain instead depends sharply on how fast influence decays with
distance, switching between weak and strong cooperation as the decay rate crosses a threshold.
Vigilance range and alignment act jointly: extending vigilance range promotes cooperation only where
the monitoring and game structures are aligned, and neither factor alone is sufficient.
Our results strongly suggest that even modest expansions of social awareness---such as
those enabled by online social platforms---can substantially reshape the landscape of
cooperative behavior in human populations.
\end{abstract}

\maketitle

\section{\label{sec:intro}Introduction}

The evolution of cooperation among self-interested individuals is one of the central
questions in evolutionary dynamics and social science~\cite{Nowak1992,Santos2005,Szabo2007,Perc2017}.
A key insight of the past two decades is that the \textit{structure} of social interactions
shapes this process profoundly: cooperation fares very differently in heterogeneous contact
networks than in well-mixed populations, and its fate changes further when one accounts for
the fact that individuals simultaneously belong to multiple social layers---family ties,
professional acquaintances, online contacts---naturally modeled as multiplex
networks~\cite{Boccaletti2014,Kivela2014}.
More recently, the recognition that many social interactions involve simultaneous group
effects---not just pairwise contacts---has motivated a rapidly growing body of work on
higher-order structures such as hypergraphs and simplicial complexes, which reveal
cooperative mechanisms invisible in dyadic models~\cite{Battiston2020,Battiston2021,Battiston2025}.
Our work is driven by a related intuition: the social signal that sustains vigilance is not
generated by a single neighbor but accumulates over a wider neighborhood, and understanding
how far that signal reaches is a step toward a richer, multi-hop picture of social pressure.
Among the mechanisms that can sustain cooperation in such structured populations,
\textit{social monitoring}---the awareness of being observed by others---occupies a special
place: unlike direct or indirect reciprocity, which depend on the history of past
interactions, monitoring acts on the instantaneous payoff structure, reducing the temptation
to defect for an individual who perceives that vigilant neighbors are watching.
The most established formalization of ``being observed'' in the indirect-reciprocity
literature is image scoring~\cite{NowakSigmund1998}, in which each
individual carries a public score that is raised or lowered by their past cooperative or
defective acts, and potential partners condition their willingness to cooperate on that
score.
Our mechanism shares with image scoring the basic idea of a signal about observed behavior
that propagates through the social network and shapes the incentive to cooperate, but
differs from it in two respects: vigilance is memoryless, depending only on the current
strategies of one's neighbors rather than an accumulated record of past play, and it acts
directly on the temptation to defect rather than through partner choice or conditional
cooperation.
Reputation has also been studied on explicit network structures: Giardini and
Vilone~\cite{GiardiniVilone2016} model gossip-based reputation spreading among agents
connected through shared group membership on a bipartite network, but restrict information
exchange to a single gossip step per pair per generation, without cascading further through
the network.
Our vigilance mechanism instead propagates continuously across $L$ circles of influence, with
weight decaying geometrically with network distance---a distance-resolved, multi-hop form of
social pressure that, to our knowledge, has not been previously combined with an evolutionary
game.
Beyond extending the vigilance mechanism of~\cite{Pereda2016} to multiple circles of
influence, the conceptual step this model takes is to introduce distance-resolved, multi-hop
social pressure into an evolutionary-game framework, bridging the literature on cooperation
in structured and multilayer populations with empirical and theoretical work on multi-step
social influence and complex contagion---two bodies of work that have so far developed
largely independently of one another.
This \textit{monitoring hypothesis}~\cite{Pereda2016} connects naturally to Axelrod's theory
of behavioral norms~\cite{Axelrod1986}: vigilance is a costly prosocial engagement that
propagates by threshold contagion, stabilizing when a sufficient fraction of neighbors
already participate and collapsing otherwise.

Pereda~\cite{Pereda2016} introduced a two-layer multiplex model coupling a weak Prisoner's
Dilemma (PD) on one layer with a Watts threshold cascade~\cite{Watts2002} of vigilance on
the other: cooperators who sense sufficient vigilance in their neighborhood lower the
temptation they offer to defectors, creating a feedback loop that sustains cooperation even
at high temptation values.
Threshold-based contagion of this kind---in which adoption occurs once the influence from
one's social environment exceeds a critical value---has been studied extensively in the
context of opinion formation and consensus~\cite{Deffuant2000,Hegselmann2002}, establishing
a rich phenomenology that connects the vigilance cascade to broader social dynamics.
Pereda and Vilone~\cite{Pereda2017} subsequently extended this analysis to additional
update rules, network topologies, and initial conditions, identifying the structural and
dynamical regimes in which the cooperation-promoting effect is strongest and those in which
it weakens.
In both studies, however, social influence is restricted to \textit{direct} neighbors,
i.e., the first circle of influence, leaving open the question of how the dynamics change
when the vigilance signal reaches further into the network.

There is extensive empirical evidence that social influence extends well beyond direct neighbors.
In large longitudinal cohort studies, both obesity and happiness have been shown to cluster
in social networks up to three degrees of separation---the friends of one's
friends' friends~\cite{Christakis2007,Fowler2008}---demonstrating that the width of social
influence is a meaningful quantity for behavioral outcomes, not merely a modeling convenience.
Vigilance, moreover, is a \textit{complex contagion}~\cite{Centola2007}: adoption requires
reinforcement from multiple contacts rather than a single exposure~\cite{Abella2023}, so
that access to a wider circle of vigilant neighbors can push an agent over its threshold even
when no single neighbor alone would suffice.
Online social platforms amplify this effect by making the behavior of distant contacts
directly observable~\cite{Centola2010}.
More recently, Miranda \textit{et al.}~\cite{Miranda2024} quantified the decay of
social influence with distance in a controlled laboratory experiment with 592 participants:
influence from second-circle contacts is approximately two-thirds that of direct neighbors,
and from third-circle contacts approximately one-third, with the second-circle effect
significantly stronger than the third.
Together, these findings suggest that restricting the vigilance signal to the immediate
neighborhood substantially underestimates the reach of social pressure in real social systems.

Motivated by this evidence, we extend the model of~\cite{Pereda2016} to $L$ circles of
influence ($L=1,2,3,4$), with the vigilance signal weighted by the geometric decay kernel
$\alpha_d = \lambda^{d-1}$; for $\lambda \approx 0.65$ this kernel reproduces the empirical
second-circle weight of~\cite{Miranda2024}, and at $L=1$ the model reduces exactly
to~\cite{Pereda2016}.
We adopt the Fermi update rule~\cite{Szabo2007,Roca2009}---new for this model---and
validate it against the replicator dynamics of~\cite{Pereda2016} at $L=1$ before turning
to $L>1$.
We then study how the vigilance range affects stationary cooperation in both correlated and
uncorrelated two-layer multiplex networks built on Barab\'asi--Albert and Erd\H{o}s--R\'enyi
topologies (mean degree $z=4$ and $z=16$), assess the sensitivity of the results to the
decay parameter $\lambda$, check their robustness to the network realization, population
size, Fermi noise, and update scheme, and test the robustness of the main result on a real
social network.

The paper is organized as follows.
Section~\ref{sec:model} presents the model in detail, including the multiplex structure,
the vigilance kernel, the update rule, and the simulation protocol.
Section~\ref{sec:results} reports the numerical results: Sec.~\ref{sec:REPvsFermi} validates
the Fermi rule against the replicator at $L=1$; Sec.~\ref{sec:correlated} presents the
main results for correlated multiplex networks across $L=1,\ldots,4$;
Sec.~\ref{sec:uncorrelated} analyzes the role of inter-layer correlation;
Sec.~\ref{sec:lambda} assesses sensitivity to $\lambda$; Sec.~\ref{sec:robustness} checks
robustness to the network realization, population size, Fermi noise, and update scheme; and
Sec.~\ref{sec:realnetwork} tests the robustness of these findings on a real social network.
Section~\ref{sec:conclusions} summarizes the findings and discusses their implications
for the spread of cooperative vigilance in real social networks.
Our main finding is that extending vigilance beyond direct neighbors---even to just the
second circle of influence---substantially promotes cooperation across all network topologies
studied, synthetic and real alike, consistent with the empirical evidence that social
influence reaches well beyond direct contacts~\cite{Christakis2007,Miranda2024}.

\section{\label{sec:model}Model}

\begin{figure}[t]
  \centering
  \includegraphics[width=\columnwidth]{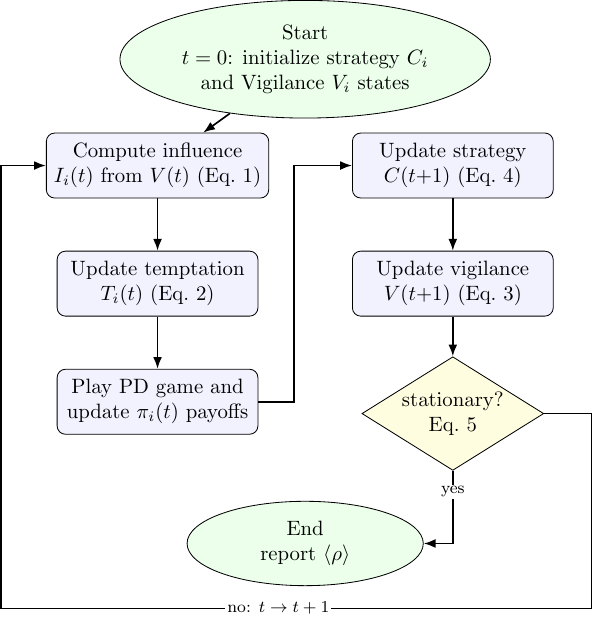}
  \caption{\label{fig:flowchart}Overview of the model ingredients and their update
    sequence within a single generation.
    Vigilance is not fixed at initialization: $V_i$ is recomputed every generation
    from the current strategy $C_i$ and the (previously computed) influence $I_i$,
    coupling the vigilance and game dynamics throughout the simulation.
    See Sec.~\ref{sec:model} for the definitions of $I_i$ (Eq.~\ref{eq:influence}),
    $T_i$ (Eq.~\ref{eq:temptation}), $V_i$ (Eq.~\ref{eq:vigilance}), the Fermi update
    (Eq.~\ref{eq:fermi}), and the stopping criterion (Eq.~\ref{eq:stopping}).}
\end{figure}

\subsection{Multiplex network and game}

We consider a population of $N=1000$ agents arranged on a two-layer multiplex network
with a shared node set: a \textit{game layer} $G_\mathrm{game}$ and a
\textit{vigilance layer} $G_\mathrm{vig}$.
In the \textit{correlated} multiplex, both layers share the same edges
($G_\mathrm{vig}=G_\mathrm{game}$); in the \textit{uncorrelated} multiplex, both layers
are independent realizations of the same random graph model.

Each agent $i$ carries two binary variables: a strategy $C_i\in\{0,1\}$
(cooperator/defector) and a vigilance state $V_i\in\{0,1\}$ (vigilant/non-vigilant).
Agents play a weak Prisoner's Dilemma on $G_\mathrm{game}$.
Payoffs are $R=1$ (cooperator--cooperator), $P=0$ (defector--defector), $S=0$
(cooperator against defector), and $T_i$ (defector against cooperator $i$), where
$T_i$ depends on the local vigilance environment of $i$ (see below).
The total payoff of agent $i$ is $\pi_i = \sum_{j\in\partial_i}[C_iC_j\cdot R +
C_i(1-C_j)\cdot S + (1-C_i)C_j\cdot T_i + (1-C_i)(1-C_j)\cdot P]$, where $\partial_i$
denotes the neighborhood of $i$ in $G_\mathrm{game}$.

The dynamics governing $T_i$, the game, and the strategy and vigilance updates are
described in the following subsections and summarized in Fig.~\ref{fig:flowchart}.

\subsection{Circles of influence and the vigilance kernel}

Let $m_i^d(t)$ and $k_i^d$ denote, respectively, the number of vigilant agents at
generation $t$ and the (fixed) total number of agents at graph distance $d$ from node
$i$ in $G_\mathrm{vig}$ (the $d$-th \textit{circle of influence}~\cite{Miranda2024}).
The vigilance influence received by agent $i$ at generation $t$, across $L$ circles, is
\begin{equation}
  I_i(t) = \min\!\left(1,\;\sum_{d=1}^{L} \lambda^{d-1}\,\frac{m_i^d(t)}{k_i^d}\right),
  \label{eq:influence}
\end{equation}
where $\lambda\in(0,1)$ is the geometric decay parameter.
For nodes whose eccentricity in $G_\mathrm{vig}$ is smaller than $L$, some shells can be
empty ($k_i^d=0$, so $m_i^d/k_i^d$ is formally undefined); we assign these terms zero
contribution to the sum rather than leave them undefined.
The kernel $\alpha_d=\lambda^{d-1}$ is intentionally \textit{not} normalized: each
additional circle adds influence on top of the previous one, so that an agent embedded
in a larger vigilant neighborhood feels stronger social pressure.
An agent exposed to $L$ circles of influence receives a stronger total vigilance signal
as $L$ increases, even when the fraction of vigilant agents is constant---converging
toward the global network state as $L$ approaches the network diameter.
This choice is deliberate, not incidental: an alternative, normalized kernel
$\alpha_d=\lambda^{d-1}/\sum_{d'=1}^L\lambda^{d'-1}$ (so that $\sum_d\alpha_d=1$) would
average, rather than accumulate, vigilance across circles.
Under the homogeneous approximation $m_i^d/k_i^d\approx v$ for all $d$---a good
approximation near the near-full-cooperation and near-full-defection outcomes typical of
this model---the normalized kernel gives $I_i^{\rm norm}=v\sum_d\alpha_d=v$, exactly
independent of $L$ and $\lambda$: extending the vigilance range would then have no effect
at all on the influence received, only redistributing a fixed total across more circles.
We confirmed this collapse both analytically and by simulation, at representative
cooperative, defective, and transition points for all four topologies: the mean
$|\langle\rho\rangle(L=1)-\langle\rho\rangle(L=4)|$ stayed below $0.01$ in every
topology (below $0.004$ overall), before adopting the unnormalized kernel used throughout
this paper, which instead lets range and magnitude increase together, consistent with the
additive weighting used
in~\cite{Miranda2024}.

The individual temptation of defecting against cooperator $i$ at generation $t$ is
\begin{equation}
  T_i(t) = 1 + (b-1)(1-I_i(t)),
  \label{eq:temptation}
\end{equation}
with $b\in[1,2]$ the baseline temptation parameter.
A fully monitored cooperator ($I_i=1$) has $T_i=1$, offering no advantage to defectors;
an unmonitored cooperator ($I_i=0$) has $T_i=b$, the standard PD temptation.
Since $I_i\leq1$, we have $T_i\geq R=1$ for all $L$, preserving the Prisoner's
Dilemma structure.
At $L=1$, Eq.~(\ref{eq:influence}) reduces to $I_i=m_i^1/k_i^1$, recovering the
model of~\cite{Pereda2016} exactly.

We choose a geometric rather than a linear decay for two reasons.
Ref.~\cite{Miranda2024} approximates the empirical influence coefficients by a linear
function of distance, $\alpha_d\propto(D-d)/(D-1)$, where $D$ is the network diameter.
This approximation, while analytically convenient for their specific experimental setting,
depends on $D$, which differs across the four topologies studied here: sparse BA and ER
networks have diameters of 6--9, while dense networks have diameters of 3--4.
Using a diameter-dependent kernel would assign different weights to the same physical
circle $d$ in different topologies, confounding the effect of vigilance range with the
effect of topology.
The geometric kernel $\alpha_d=\lambda^{d-1}$ is independent of $D$: the same $\lambda$
gives the same relative weights in all networks, so that differences in results across
topologies reflect only the network structure, not the kernel definition.
Second, $\lambda$ is a single continuous parameter that smoothly interpolates between
purely local social pressure ($\lambda=0$, equivalent to $L=1$) and uniform weighting
across all circles ($\lambda\to1$), allowing us to study the effect of the decay rate
independently of the range $L$ (Sec.~\ref{sec:lambda}).

\subsection{\label{sec:vigilance}Vigilance dynamics}

Vigilance follows a Watts threshold rule~\cite{Watts2002}, re-evaluated every
generation: agent $i$ is vigilant at generation $t+1$ if and only if it is a
cooperator at $t+1$ and the influence it received at the \emph{previous} generation
met or exceeded a threshold $\theta$,
\begin{equation}
  V_i(t+1) = C_i(t+1) \cdot \mathbf{1}[I_i(t) \geq \theta].
  \label{eq:vigilance}
\end{equation}
Vigilance is thus not fixed at initialization: $V_i(t+1)$ is recomputed every
generation from the newly updated strategy $C_i(t+1)$ (Eq.~\ref{eq:fermi}) and the
influence $I_i(t)$ already computed this generation (Eq.~\ref{eq:influence}), coupling
the vigilance and game dynamics throughout the simulation.
At $\theta=0$, all cooperators are vigilant; increasing $\theta$ demands a denser
vigilant neighborhood before an agent participates in the cascade.

A simple mean-field argument gives intuition for how the vigilance range $L$ shapes this
cascade. Under a homogeneous approximation in which the vigilant fraction in each distance
shell is approximately $v$, the geometric kernel (Eq.~\ref{eq:influence}) gives
$I(L)\approx v\sum_{d=1}^L\lambda^{d-1} = v(1-\lambda^L)/(1-\lambda)$, so that the
activation condition $I\geq\theta$ corresponds to a critical vigilant fraction
\begin{equation*}
  v_c(L) \approx \theta\,\frac{1-\lambda}{1-\lambda^L}.
\end{equation*}
Since $1-\lambda^L$ increases with $L$ (from $1-\lambda$ at $L=1$ toward $1$ as
$L\to\infty$), $v_c(L)$ decreases monotonically with $L$: a smaller fraction of vigilant
neighbors suffices to trigger the cascade as the vigilance range widens. The increment
$\lambda^{L-1}$ separating consecutive values of $1-\lambda^L$ shrinks geometrically with
$L$, so this argument also predicts a diminishing-returns pattern: most of the reduction in
$v_c$ should already be achieved by $L=2$, with each additional circle contributing
progressively less.

\subsection{Strategy update}

Strategies are updated synchronously using the Fermi rule~\cite{Szabo2007,Roca2009}.
At each generation $t$, every agent $i$ selects a random neighbor $j$ in $G_\mathrm{game}$
and adopts $j$'s strategy at $t+1$, i.e.\ sets $C_i(t+1)=C_j(t)$, with probability
\begin{equation}
  p(i\leftarrow j) = \frac{1}{1+\exp[(\pi_i(t)-\pi_j(t))/K]},
  \label{eq:fermi}
\end{equation}
with noise parameter $K=0.1$ (corresponding to an inverse temperature $\beta=1/K=10$
in the notation of~\cite{Roca2009}).
This rule is novel for the present model; \cite{Pereda2017} compared replicator,
unconditional imitation, and mixed rules but did not include Fermi.
We validate the Fermi rule against the replicator dynamics of~\cite{Pereda2016}
at $L=1$ in Sec.~\ref{sec:REPvsFermi}.

\subsection{Network topologies and initial conditions}

We consider Barab\'asi--Albert (BA)~\cite{Barabasi1999} and
Erd\H{o}s--R\'enyi (ER)~\cite{Erdos1959} random graphs with $N=1000$ nodes and
mean degree $z\in\{4,16\}$.
BA networks are generated by preferential attachment with $m=z/2$ edges added per
new node; ER networks with connection probability $p=z/(N-1)$.
In both cases the largest connected component is retained.
A single network realization is fixed per topology and degree; 100 independent
initial conditions are used to compute ensemble averages.
Initial conditions assign cooperator/defector uniformly at random (50\% each), with
each cooperator independently vigilant with probability 1/2, so that 25\% of the
population starts as vigilant cooperators, matching~\cite{Pereda2016}.

\subsection{Stopping criterion}

Simulations run until the stationary state is detected adaptively.
We partition the trajectory into non-overlapping windows of 100 generations and, after a
minimum transient of 500 generations, compare the mean cooperator fraction of consecutive
windows,
\begin{equation}
  W_k = \frac{1}{100}\sum_{t=100(k-1)}^{100k-1} \langle\rho\rangle_t,
  \qquad
  |W_k - W_{k-1}| < 10^{-2},
  \label{eq:stopping}
\end{equation}
where $\langle\rho\rangle_t$ is the mean cooperator fraction at generation $t$ and $W_k$ is
the mean over window $k$.
When Eq.~(\ref{eq:stopping}) is satisfied, $W_k$ is recorded as the stationary value.
If convergence is not reached within $5\times10^5$ generations, the mean of the final
window is recorded instead.
This criterion serves the same purpose as that of~\cite{Pereda2016} but is adaptive:
it does not require a fixed transient length, which would be computationally prohibitive
at $L>1$ because the vigilance shells can encompass a large fraction of the network.

\subsection{Parameter summary}

We sweep $b\in\{1.0,1.1,\ldots,2.0\}$ (11 values) and $\theta\in\{0.0,0.1,\ldots,1.0\}$
(11 values), for all combinations of $L\in\{1,2,3,4\}$, network topology (BA, ER),
mean degree ($z=4$, $z=16$), and multiplex correlation (correlated, uncorrelated).
The default decay parameter is $\lambda=0.5$; sensitivity to $\lambda$ is studied
in Sec.~\ref{sec:lambda} for $\lambda\in\{0.1,0.25,0.5,0.75,0.9\}$.
All parameter combinations are summarized in Table~\ref{tab:params}.

\begin{table}[t]
\caption{\label{tab:params}Summary of simulation parameters.}
\begin{ruledtabular}
\begin{tabular}{ll}
Parameter & Values \\
\hline
Population size $N$ & 1000 \\
Replications & 100 \\
Temptation $b$ & $1.0, 1.1, \ldots, 2.0$ (11 values) \\
Threshold $\theta$ & $0.0, 0.1, \ldots, 1.0$ (11 values) \\
Circles of influence $L$ & 1, 2, 3, 4 \\
Decay parameter $\lambda$ (default) & 0.5 \\
Decay parameter $\lambda$ (sensitivity) & 0.1, 0.25, 0.5, 0.75, 0.9 \\
Fermi noise $K$ & 0.1 \\
Network topologies & BA, ER \\
Mean degree $z$ & 4, 16 \\
Multiplex correlation & correlated, uncorrelated \\
\end{tabular}
\end{ruledtabular}
\end{table}

\subsection{Computational implementation}

Simulations were implemented in Python using just-in-time compiled kernels
(\texttt{numba}) for the inner loops (influence computation, payoff accumulation,
and Fermi update), and \texttt{joblib} thread-based parallelism over the 100
independent replications.
Vigilance shells (the sets of nodes at each graph distance $d=1,\ldots,L$) were
precomputed via breadth-first search and stored in compressed sparse-row format,
so that shell construction is performed once per network and reused across all
parameter combinations.
All code and simulation outputs, together with notes on typical runtimes, are
publicly available at \url{https://github.com/mpereda/long-range-social-pressure}.

\section{\label{sec:results}Results}

The results below use a single fixed network realization per topology and degree,
$N=1000$, Fermi noise $K=0.1$, and synchronous strategy updating throughout, as detailed
in Sec.~\ref{sec:model}; Sec.~\ref{sec:robustness} checks that none of these choices
drives our conclusions.

\subsection{\label{sec:REPvsFermi}Replicator vs.\ Fermi at \texorpdfstring{$L=1$}{L=1}}

Before studying the effect of long-range vigilance, we check that the cooperation
results at $L=1$ are robust to the choice of update rule by comparing the Fermi rule
with the replicator dynamics of~\cite{Pereda2016} (see Supplemental Material,
Sec.~S1, for the derivation of this rule and its relation to mean-field
replicator dynamics).
At $L=1$, Eq.~(\ref{eq:influence}) reduces to $I_i = m_i^1/k_i^1$, recovering the
original model of~\cite{Pereda2016} exactly.

\begin{figure}[t]
  \centering
  \includegraphics[width=\columnwidth]{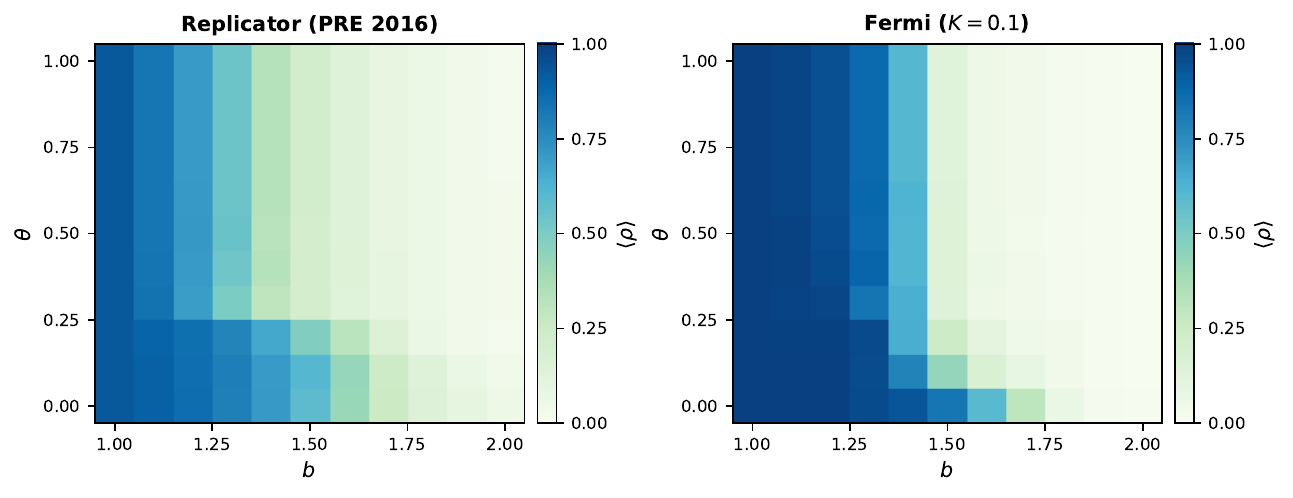}
  \caption{Stationary cooperation fraction $\langle\rho\rangle$ as a function of
    temptation $b$ and vigilance threshold $\theta$ at $L=1$ for BA $z=16$:
    replicator (left) and Fermi $K=0.1$ (right).
    Both rules produce qualitatively identical phase diagrams; Fermi sustains more
    cooperation at low-to-intermediate $b$ and the replicator at higher $b$, crossing over
    near $b\approx1.4$--$1.5$ (discussed in the text).
    The high variance across replications near the transition region is discussed in the
    text; it is not visible in these mean-value heatmaps.
    Full heatmaps for all four topologies are in Fig.~S2.
    $N=1000$, 100 replications.}
  \label{fig:REPvsFermi}
\end{figure}

Figure~\ref{fig:REPvsFermi} shows the stationary cooperation fraction $\langle\rho\rangle$
as a function of $b$ and $\theta$ for both update rules in BA $z=16$;
results for all four topologies are provided in the Supplemental Material (Fig.~S2).
The two rules produce qualitatively identical phase diagrams across all networks (Fig.~S2),
though the size and pattern of the quantitative differences between them depend strongly on
topology and mean degree.
BA $z=16$ is the topology where the two rules differ most on average (mean
$|\langle\rho\rangle_\mathrm{rep}-\langle\rho\rangle_\mathrm{fermi}|=0.13$, versus $0.09$
for BA $z=4$, $0.07$ for ER $z=4$, and $0.10$ for ER $z=16$), and the difference there is
systematic rather than confined to a narrow region: the Fermi rule sustains more
cooperation at low-to-intermediate temptation ($b\lesssim1.4$), while the replicator
sustains more at higher temptation ($b\gtrsim1.5$), with a crossover near $b\approx1.4$--$1.5$
(Fig.~S2).
In BA $z=4$, by contrast, the Fermi rule yields higher cooperation across the entire
sampled range, with the gap growing monotonically from $b=1.0$ to $b=1.9$ and no crossover.
In both Erd\H{o}s--R\'enyi topologies, cooperation collapses from near-full to near-zero
over a comparatively narrow range of $b$ --- we refer to this range as the topology's
\emph{collapse boundary} throughout this section.
The two rules agree closely away from it (mean $|\Delta\langle\rho\rangle|\leq0.03$ at the
lowest and highest sampled $b$ values) but diverge substantially within it, and the width
of that disagreement tracks how sharp each topology's own collapse boundary is.
For ER $z=16$, whose collapse is confined to a narrow range of $b$ near $1.0$--$1.2$, the
disagreement between rules is correspondingly a narrow, localized spike: a small offset
between the two rules' critical temptations is amplified by the steepness of the transition
into a large discrepancy (mean $|\Delta\langle\rho\rangle|$ up to $0.35$, single-cell peak
$\approx0.86$) confined to $b\approx1.1$--$1.2$ and falling below $0.03$ by $b=1.7$.
For ER $z=4$, whose transition is more gradual, the region of disagreement is
correspondingly broader, spanning roughly $b=1.2$--$1.8$ (mean $|\Delta\langle\rho\rangle|$
up to $0.15$ at $b=1.3$) --- smaller in peak magnitude than ER $z=16$'s spike, but spread
over more of the sampled range.
This pattern is consistent with the known sensitivity of cooperation-transition boundaries
to the choice of update rule~\cite{Roca2009}: a small offset between the two rules'
critical temptations is unavoidable, and how much of parameter space it disrupts depends on
how sharp the underlying transition is --- sharpest, and therefore narrowest in its
disruption, for ER $z=16$; broadest for BA $z=16$ (discussed above), whose gradual
transition instead lets the two rules' distinct boundaries occupy a genuinely shared,
extended region of parameter space.

We also note that BA $z=16$ shows the highest variance across replications of any
topology: standard deviations reach $\sigma\approx0.47$ near the transition region
($b=1.6$, $\theta=0$), compared to $\sigma\lesssim0.14$ for BA $z=4$; ER networks also
show substantial variance in places ($\sigma$ up to $0.35$ for ER $z=4$ at $b=1.6$ and
$0.30$ for ER $z=16$ at $b=1.1$) --- in each case at their own collapse boundary, the same
region where the two update rules disagree most (above).
For BA $z=16$, this elevated variance reflects a bistable regime in which the outcome
depends on the initial condition: replications that start with cooperative hubs converge
to high cooperation, while those starting with defecting hubs collapse to defection,
producing a bimodal distribution of final states rather than a well-defined stationary
value (Fig.~S3).
At the point of highest variance, $57.5\%$ of 200 replications converge to the cooperation
attractor ($\langle\rho\rangle>0.8$) and $41.5\%$ to the defection attractor
($\langle\rho\rangle<0.2$), with almost none in between.
The two attractors are not symmetric: the cooperation attractor is a sharp, narrow peak at
$\langle\rho\rangle\approx1$, while the defection attractor is a broader plateau spanning
$\langle\rho\rangle\approx0$--$0.2$ (Fig.~S3).
This follows directly from the vigilance rule (Eq.~\ref{eq:vigilance}): only cooperators
can be vigilant, so the positive feedback between cooperation and vigilance that locks the
system tightly onto the $\langle\rho\rangle\approx1$ attractor has no counterpart that
similarly locks it onto $\langle\rho\rangle\approx0$, where the dynamics instead reduce to
an ordinary Prisoner's Dilemma with no monitoring effect (Supplemental Material, Sec.~S3).
The same asymmetry is why the histogram bars near $\langle\rho\rangle\approx0$ are individually
shorter than the single bar at $\langle\rho\rangle\approx1$ despite the comparable total share
of replications on each side: that share is spread across several bins rather than
concentrated in one (Supplemental Material, Sec.~S3).
This behavior, already present at $L=1$, will be relevant when interpreting the high
variance observed for BA $z=16$ in the $L>1$ results of the following sections.

The qualitative agreement between the two rules at $L=1$ --- the same phase-diagram shape
and collapse-boundary structure, despite the topology-dependent quantitative differences
detailed above --- validates our implementation against~\cite{Pereda2016}.
We adopt the Fermi rule for all $L>1$ results on this basis; we did not repeat the $L=1$
validation exercise for each $L>1$ setting (Supplemental Material, Sec.~S1).

\subsection{\label{sec:correlated}Long-range vigilance in the correlated multiplex}

Having established the update-rule robustness at $L=1$, we now study how extending
the vigilance range to $L>1$ circles affects stationary cooperation in the correlated
multiplex, where both layers share the same network realization.

\begin{figure}[t]
  \centering
  \includegraphics[width=\columnwidth]{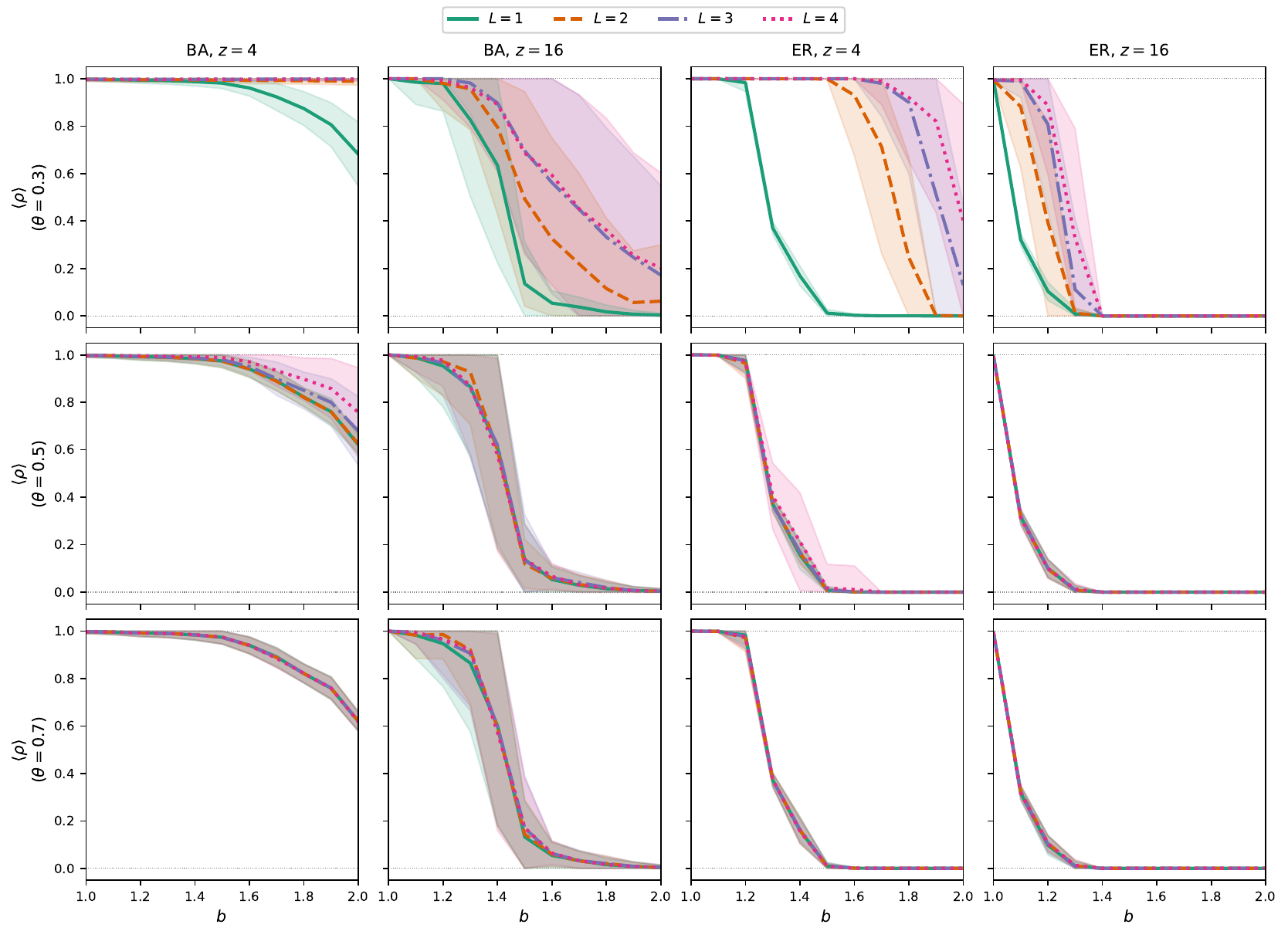}
  \caption{Stationary cooperation fraction $\langle\rho\rangle$ versus temptation $b$
    at fixed $\theta=0.3$ (top), $0.5$ (middle), and $0.7$ (bottom) in the correlated
    multiplex (Fermi rule, $\lambda=0.5$).
    Columns correspond to (a) BA $z=4$, (b) BA $z=16$, (c) ER $z=4$, (d) ER $z=16$.
    Curves: $L=1$ (teal, solid), $L=2$ (orange, dashed), $L=3$ (purple, dash-dot),
    $L=4$ (magenta, dotted).
    Shaded bands indicate $\pm1$ standard deviation across 100 replications.
    $N=1000$.}
  \label{fig:correlated-marginals}
\end{figure}

Figure~\ref{fig:correlated-marginals} shows the stationary cooperation fraction
$\langle\rho\rangle$ as a function of temptation $b$ at three representative threshold
values ($\theta=0.3,0.5,0.7$) for $L=1,2,3,4$ in the correlated multiplex.
Full heatmaps of $\langle\rho\rangle(b,\theta)$ for all four network topologies are
provided in the Supplemental Material (Fig.~S4).
Extending the vigilance range promotes cooperation in all cases: the critical temptation
above which cooperation collapses shifts to higher $b$ values as $L$ increases, with the
largest gains concentrated at low-to-intermediate threshold ($\theta \lesssim 0.5$).

The effect is most pronounced and cleanest in sparse Erd\H{o}s--R\'enyi networks ($z=4$).
At $\theta=0.3$, cooperation collapses sharply to $\langle\rho\rangle\approx0$ for
$b\gtrsim1.3$ at $L=1$; extending to $L=2$ raises this critical temptation to
$b\approx1.7$, and $L=3$ pushes it to $b\approx1.9$, while at $L=4$ cooperation
remains near $\langle\rho\rangle\approx0.40$ even at $b=2.0$
[Fig.~\ref{fig:correlated-marginals}(c), top row].
This corresponds to a near-complete recovery of the cooperation-sustaining region.

In sparse Barab\'asi--Albert networks ($z=4$), the effect is equally strong but takes a
different form: cooperation never collapses to zero at $L=1$ in this topology, instead
declining smoothly from $\langle\rho\rangle\approx0.998$ at $b=1.0$ to $0.681$ at
$b=2.0$ for $\theta=0.3$.
Already at $L=2$, cooperation is essentially full across the entire parameter space
($\langle\rho\rangle\geq0.988$ for all $b$ at $\theta=0.3$), with $L=3$ and $L=4$ adding
only marginal improvements.
The preferential-attachment structure of BA networks, with its highly connected hubs,
provides an inherently stronger substrate for vigilance propagation, so that even one
additional circle suffices to lock the system into the cooperative attractor.

Dense networks ($z=16$) behave differently.
In ER $z=16$, the cooperative region at $L=1$ is confined to a narrow strip near $b=1.0$,
and $L>1$ expands it modestly: at $\theta=0.3$ the critical temptation shifts from
$b\approx1.1$ to $b\approx1.2$--$1.3$ for $L=4$, but cooperation still collapses by
$b=1.4$ regardless of $L$.
In BA $z=16$, cooperation at $L=1$ is substantially higher than in ER $z=16$ due to the
heterogeneous degree distribution, and each additional circle provides a measurable but
diminishing benefit; however, the variance across replications is markedly larger than in
the $z=4$ case [Fig.~\ref{fig:correlated-marginals}(b), top row], with standard
deviations reaching $\sigma\approx0.47$ near the transition region.
This high variance is characteristic of systems with multiple coexisting attractors,
consistent with the high variance already observed at $L=1$ in this topology (Sec.~\ref{sec:REPvsFermi}).

\begin{figure}[t]
  \centering
  \includegraphics[width=\columnwidth]{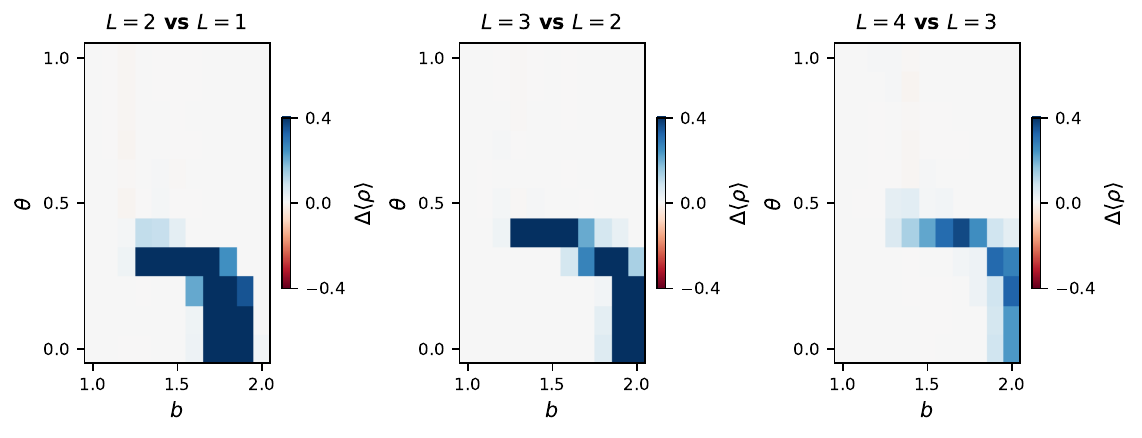}
  \caption{Marginal cooperation gain $\langle\rho\rangle(L) - \langle\rho\rangle(L-1)$
    contributed by each additional circle, for $L=2$ vs.\ $1$ (left), $3$ vs.\ $2$
    (center), and $4$ vs.\ $3$ (right) in ER $z=4$, correlated multiplex ($\lambda=0.5$).
    Blue indicates an increase in cooperation relative to the previous $L$, red a
    decrease; the region of largest marginal gain shrinks and shifts to higher $b$ as $L$
    increases.
    Marginal difference maps for all topologies are in Fig.~S5; the cumulative
    (baseline-referenced) version is in Fig.~S6.}
  \label{fig:correlated-diff}
\end{figure}

The benefit that a wider vigilance range provides can be read off in two complementary
ways: how much each additional circle contributes on its own --- the marginal gain,
Fig.~\ref{fig:correlated-diff} (Fig.~S5 for all topologies) --- and how much of the total
achievable gain has accumulated by a given $L$ --- the cumulative gain, Fig.~S7.

Figure~\ref{fig:correlated-diff} shows the \emph{marginal} cooperation gain
$\langle\rho\rangle(L) - \langle\rho\rangle(L-1)$ contributed by each additional circle,
for ER $z=4$, the topology with the largest and most regular gain; the marginal maps for
all four topologies are in Fig.~S5, alongside the cumulative (baseline-referenced)
version of the same comparison in Fig.~S6.
The marginal maps reveal three consistent patterns.
First, the marginal gain shrinks with $L$: in ER $z=4$, its peak value falls from
$0.99$ ($L=2$ vs.\ $L=1$) to $0.65$ ($L=3$ vs.\ $L=2$) to $0.36$ ($L=4$ vs.\
$L=3$), consistent with the geometric decay $\lambda^{d-1}=0.5^{d-1}$ that assigns
weights $1$, $0.5$, $0.25$ to circles $d=1,2,3$; the same decay is visible in BA $z=4$,
where the peak marginal gain falls from $0.31$ to $0.17$ to $0.08$ across the same three
steps --- smaller in absolute terms, since BA $z=4$ already cooperates well at $L=1$, but
following essentially the same geometric ratio.
Second, the region where each additional circle helps most does not simply shrink in
place: it visibly shifts to higher $b$ as $L$ increases (peaking near $b\approx1.5$ for
$L=2$, $b\approx1.8$ for $L=3$, and $b\approx1.7$ for $L=4$ in ER $z=4$), since the
lower-$b$ region is already saturated by the earlier circles.
Third, in BA $z=16$ (Fig.~S5) a small number of $(b,\theta)$ grid cells show a slightly
negative marginal gain at $L=4$ (differences of at most $0.07$), within the range
attributable to the high variance in that regime; no systematic reduction was observed
for $L=2$ or $L=3$.

A different, cumulative question is what fraction of the total $L=1\to4$ gain is
already achieved by a given $L$, and here the answer is markedly topology dependent
(Supplemental Material, Fig.~S7): in both Erd\H{o}s--R\'enyi densities, the $L=1\to2$
step alone already captures the large majority of the total gain (93--96\%), and the
further step to $L=3$ adds only $4$--$8\%$ as much again; in both Barab\'asi--Albert
densities the cumulative gain instead builds much more evenly across the first two
additional circles ($52$--$65\%$ already reached at $L=2$, a further increment
$59$--$87\%$ as large still to come at $L=3$). Sparse Erd\H{o}s--R\'enyi topologies thus
front-load nearly the entire benefit into the first additional circle, while
Barab\'asi--Albert topologies spread it across the first two, converging by $L=3$ to
essentially the same outcome across all four topologies ($96$--$103\%$ of the total gain
captured).

This diminishing-returns pattern can be compared against the mean-field prediction of
Sec.~\ref{sec:vigilance}, $v_c(L)\approx\theta(1-\lambda)/(1-\lambda^L)$: at $\lambda=0.5$, the
theory predicts that $71\%$ of the total reduction in the critical vigilant fraction from
$L=1$ to $L=4$ is already achieved at $L=2$, and $92\%$ at $L=3$.
Measuring the analogous quantity directly from the critical temptation $b_c(L)$ at
$\theta=0.3$ (the value used throughout this subsection), the fraction of the total
$b_c(1)\to b_c(4)$ shift captured at $L=2$ is $67\%$ for ER $z=4$, $54\%$ for ER $z=16$,
and $30\%$ for BA $z=16$; at $L=3$ it is $89\%$, $87\%$, and $95\%$, respectively.
The theory's $L=3$ prediction ($92\%$) matches the data well across all three topologies;
its $L=2$ prediction ($71\%$) is close for ER $z=4$ but overestimates the empirical share
for ER $z=16$ and especially BA $z=16$, where hub-driven heterogeneity is strongest.
The homogeneous mean-field approximation therefore captures the correct qualitative
trend---and roughly where saturation sets in---but underestimates how gradual the real,
degree-heterogeneous network's response is at low $L$.
This is corroborated by an independent metric: Barab\'asi--Albert also lags
Erd\H{o}s--R\'enyi in the fraction of the total $L=1\to4$ cooperation gain captured at
$L=2$ (see above; Supplemental Fig.~S7).

\subsection{\label{sec:uncorrelated}Role of inter-layer correlation}

The results so far used a correlated multiplex in which both layers share the same
edges, so that an agent's social monitors are precisely the same individuals it plays
the game against and learns from.
This alignment is not guaranteed in real social systems: the people who observe your
behavior online are not necessarily your colleagues, neighbors, or competitors; your
professional contacts may monitor you in ways that are entirely uncorrelated with
who you actually cooperate or compete with in a given context.

\begin{figure}[t]
  \centering
  \includegraphics[width=\columnwidth]{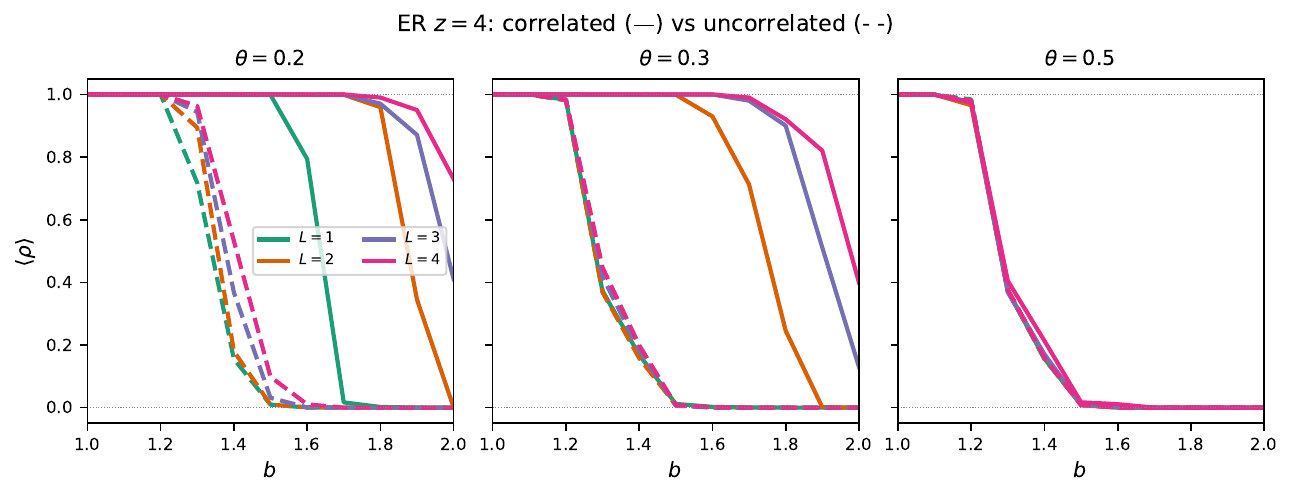}
  \caption{Stationary cooperation fraction $\langle\rho\rangle$ versus temptation $b$ for
    ER $z=4$ at $\theta=0.2,0.3,0.5$: correlated (solid) versus uncorrelated (dashed)
    multiplex, for $L=1$ (teal), $2$ (orange), $3$ (purple), $4$ (magenta).
    The gap between correlated and uncorrelated curves grows with $L$ near the
    cooperation transition (left and center panels); at $\theta=0.5$ (right panel)
    both configurations yield comparable outcomes across the full range of $b$ shown.
    Full comparison for all topologies in Fig.~S8.}
  \label{fig:uncorrelated}
\end{figure}

To isolate the role of this alignment, we now compare the correlated multiplex of
Sec.~\ref{sec:correlated} with an \emph{uncorrelated} multiplex in which
$G_\mathrm{vig}$ is an independent realization of the same random-graph model as
$G_\mathrm{game}$.
In the uncorrelated case, being monitored by vigilant agents still reduces an
individual's temptation parameter $T_i$---because $I_i$ is computed from
$G_\mathrm{vig}$ regardless of who one's game partners are---but the protective
effect is no longer spatially coherent: a cluster of cooperative, vigilant agents
in $G_\mathrm{vig}$ does not correspond to the same cluster of individuals actually
interacting in the game, so the mutual reinforcement that makes the cooperative
cluster evolutionarily stable in the correlated case is broken.
Figure~\ref{fig:uncorrelated} compares correlated and uncorrelated stationary cooperation
for ER $z=4$, the topology where the effect of correlation is largest; results for all four
topologies are in Fig.~S8.

As Fig.~\ref{fig:uncorrelated} shows, inter-layer correlation matters, and its effect
grows with $L$, most prominently in sparse Erd\H{o}s--R\'enyi networks.
In ER $z=4$, decorrelating the layers increasingly suppresses the benefit of extending
$L$: the correlated critical temptation at $\theta=0.2$ shifts markedly upward as $L$
increases---reaching $b>2.0$ at $L=4$, as shown in Sec.~\ref{sec:correlated}---while
the uncorrelated critical temptation remains essentially fixed near $b\approx1.3$--$1.5$
for every $L$.
At $b=2.0$, $\theta=0.2$, this amounts to $\langle\rho\rangle\approx0.73$ in the correlated
case versus $\langle\rho\rangle\approx0$ in the uncorrelated case at $L=4$.
Averaged over the full $(b,\theta)$ region where the effect is visible, the mean
$|\langle\rho\rangle_\mathrm{uncorr}-\langle\rho\rangle_\mathrm{corr}|$ grows from $0.06$
at $L=1$ to $0.22$ at $L=4$ in ER $z=4$---more than a threefold increase---indicating that
extending the vigilance range amplifies, rather than dilutes, the importance of layer
correlation.
ER $z=16$ shows the same sign of effect but confined to a narrow region near the
collapse boundary (e.g., at $\theta=0.1$, $b=1.3$, $L=4$: $\langle\rho\rangle\approx0.75$
correlated versus $0.01$ uncorrelated, a point with high variance across replications,
$\sigma\approx0.41$, characteristic of a transition region), and the mean difference stays
below $0.03$ across all $L$.

Barab\'asi--Albert networks are far less sensitive to inter-layer correlation than
Erd\H{o}s--R\'enyi at either density, though not entirely insensitive: the mean difference
stays at or below $0.011$ throughout for $z=4$, and at or below $0.021$ for $z=16$ (rising
from $0.013$ at $L=1$ before leveling off), both roughly an order of magnitude smaller than
the corresponding ER $z=4$ values. The critical temptation in BA $z=16$ shifts similarly
whether the layers are correlated or not at $L=1$ (differences of at most $0.03$ in $b$
across $\theta$), but a small residual gap opens up by $L=4$ (up to $0.10$ in $b$). The
inter-layer degree correlation quantified below does not itself depend on $L$, so this
growing gap means the residual misalignment it reflects compounds as $L$ grows---the same
pattern responsible for the much larger correlation effect in ER, just far weaker in BA.
Part of this robustness reflects a construction property rather than a purely dynamical
one. Although $G_\mathrm{vig}$ and $G_\mathrm{game}$ are independent realizations of the
same random-graph model, node indices in the standard Barab\'asi--Albert construction are
assigned in arrival order, and preferential attachment consistently favors low-index,
early-arriving nodes regardless of the random seed; the same node is therefore likely to
be a hub in both independently generated layers.
We quantify this with the inter-layer degree correlation
$r_k=\mathrm{corr}(k_i^\mathrm{game},k_i^\mathrm{vig})$: $r_k=0.68$ for $z=4$ and $0.86$
for $z=16$ in the nominally uncorrelated BA multiplex, compared with $r_k\approx0$ for
Erd\H{o}s--R\'enyi networks, whose nodes are exchangeable and carry no such arrival-order
signal.
The high-degree hubs that dominate vigilance propagation in BA networks thus do tend to
occupy structurally similar roles in $G_\mathrm{vig}$ and $G_\mathrm{game}$ even when the
two layers share no edges---but this alignment follows substantially from the shared
node-arrival order of the network generator, not from a property intrinsic to the
vigilance dynamics itself, and the residual sensitivity BA does show to decorrelation,
reported above, is consistent with $r_k$ falling short of 1 rather than reaching it.

To test this directly, we additionally randomly permuted $G_\mathrm{vig}$'s node labels
in the uncorrelated BA multiplex, destroying the residual degree correlation
($r_k\to0$) while leaving edge overlap and $G_\mathrm{vig}$'s own topology unchanged.
At the representative points where the correlated-to-uncorrelated change is itself
non-negligible ($6$ of the $14$ points tested), shuffling further reduces cooperation
in the same direction in most cases ($4$ of $6$), consistent with $r_k$ contributing to
BA's residual sensitivity; the additional shift is small, however, and does not reach
statistical significance at the $100$-replication sample size used here (full details
in Supplemental Material, Sec.~S9).

The contrast between ER and BA points to the mechanism responsible for the correlation
effect.
When $G_\mathrm{vig}=G_\mathrm{game}$, the agents whose vigilance reduces an individual's
temptation to defect are the same agents that individual interacts with in the game; a
locally reinforcing cluster of vigilant cooperators is therefore also a cluster of mutually
protected game partners.
When the layers are uncorrelated, this alignment is broken: an agent may receive strong
vigilance influence from a wide vigilance neighborhood while its actual game opponents
draw their own vigilance from an unrelated, possibly far less vigilant, part of the network.
Extending $L$ widens the vigilance neighborhood in both cases, but only in the correlated
multiplex do those additional monitors coincide with actual game partners; in the
uncorrelated multiplex the wider reach covers game-irrelevant contacts, which is why the
critical temptation does not advance with $L$.

We also checked whether the correlated-to-uncorrelated change is a smooth crossover or an
abrupt jump as a function of how aligned the two layers are. Starting from $G_\mathrm{vig}=
G_\mathrm{game}$ and applying degree-preserving edge rewiring (ER $z=4$, the topology most
sensitive to correlation), the population-averaged $\langle\rho\rangle$ decreases smoothly
and monotonically as edge overlap is reduced, at every $b$ tested, sharpening as $b$
approaches the collapse boundary. This smooth average, however, is built from individual
network replications that do not decline gradually: right at the crossover, the
standard deviation across replications spikes sharply (up to $0.5$), the same
bistable-switching signature already documented in Sec.~\ref{sec:REPvsFermi} (Fig.~S3):
individual replications settle into either near-full cooperation or near-full defection,
and it is the changing balance between the two, not a gradual per-replication decline, that
produces the smooth population average.
Full details in Supplemental Material, Sec.~S10.

\subsection{\label{sec:lambda}Sensitivity to the decay parameter \texorpdfstring{$\lambda$}{lambda}}

The results above use $\lambda=0.5$ as an illustrative default; we now examine how
sensitive the cooperation gain from extending $L$ is to this choice.
We sweep $\lambda\in\{0.1,0.25,0.5,0.75,0.9\}$ for $L=2,3,4$ in the correlated multiplex,
focusing on Barab\'asi--Albert networks at $z=4$ and $z=16$, where network density
produced the sharpest qualitative contrast in the correlated-multiplex results
(Sec.~\ref{sec:correlated}).

Figures~S12--S14 show the full $(b,\theta)$ heatmaps for all five $\lambda$ values at
$L=2$, $L=3$, and $L=4$ respectively.
In sparse BA networks ($z=4$), $\lambda$ raises cooperation smoothly and modestly at
every $L$: the grid-averaged $\langle\rho\rangle$ gains only $0.05$ across the full
$\lambda$ range at $L=4$, with comparable or smaller gains at $L=2$ and $L=3$.
Dense BA networks ($z=16$) behave qualitatively differently at every $L$: cooperation
grows modestly for $\lambda\leq0.5$ and then jumps sharply between $\lambda=0.5$ and
$\lambda=0.75$, a pattern visible in Figs.~S12 and~S13 and most pronounced at $L=4$
(Fig.~S14).
The total grid-averaged gain across the full $\lambda$ range grows with $L$: $0.14$ at
$L=2$, $0.31$ at $L=3$, and $0.35$ at $L=4$, because each additional circle compounds
the weight $\lambda^{d-1}$ assigned to the already-influential hubs.

\begin{figure}[tp]
  \centering
  \includegraphics[width=\columnwidth]{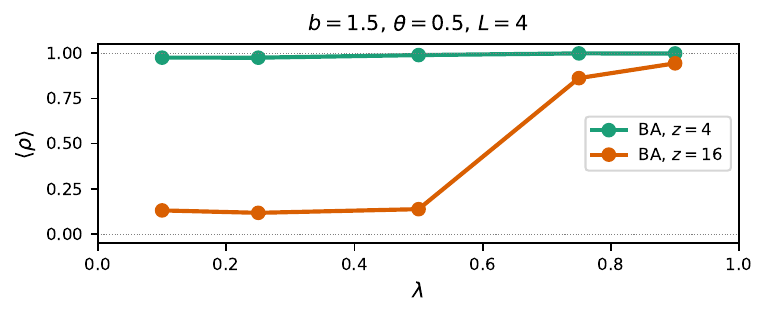}
  \caption{Stationary cooperation fraction $\langle\rho\rangle$ versus decay parameter
    $\lambda$ at $b=1.5$, $\theta=0.5$, $L=4$, correlated multiplex, Fermi rule.
    BA $z=4$ (teal) increases smoothly with $\lambda$; BA $z=16$ (orange) is essentially
    flat for $\lambda\leq0.5$ and jumps sharply between $\lambda=0.5$ and $\lambda=0.75$.
    Full $(b,\theta)$ heatmaps for both networks and all five $\lambda$ values in
    Fig.~S14.}
  \label{fig:lambda}
\end{figure}

Figure~\ref{fig:lambda} shows the cooperation fraction at the midpoint of both scanned
ranges ($b=1.5$, $\theta=0.5$) as a function of $\lambda$.
At $L=4$, the transition is starkest at this point: $\langle\rho\rangle$ moves from
$0.12$--$0.14$ for $\lambda\leq0.5$ to $0.86$ at $\lambda=0.75$ and $0.94$ at
$\lambda=0.9$, a jump of more than $0.7$ within a single step of the $\lambda$ grid.
The standard deviation mirrors this jump, rising from $\sigma\approx0.10$--$0.15$ below
the transition to $\sigma=0.33$ at $\lambda=0.75$---the signature of a system crossing
between two coexisting attractors, consistent with the bistability already observed at
$L=1$ in this same topology (Sec.~\ref{sec:REPvsFermi}).
This places the dense, hub-dominated topology in a qualitatively different regime from the
sparse one with respect to $\lambda$: where sparse networks respond gradually to the decay
rate, dense networks with strong hubs can switch abruptly between a low- and a
high-cooperation regime as $\lambda$ crosses a threshold between $0.5$ and $0.75$.

\subsection{\label{sec:robustness}Robustness to network realization, population size, and update scheme}

The results above use a single fixed network realization per topology and degree,
$N=1000$, Fermi noise $K=0.1$, and synchronous strategy updating throughout
(Sec.~\ref{sec:model}).
Given the sharp transitions and bistable behavior reported above, we checked that none of
these choices drives our conclusions.

First, we checked whether the cooperation gain from extending $L$ holds up across
independently generated networks. We reran the full $L=1,2,3,4$ range on 30 independently
generated network realizations, at the same 100-replication budget used throughout the
paper, for two $(b,\theta)$ points per topology with the largest $L=1\to4$ gain along the
$\theta=0.3$ reference line already used for the critical-temptation percentages
(Sec.~\ref{sec:correlated}) --- 8 points in total. The gain never changes sign in 239 of the 240 (point,
realization) combinations, and the population-averaged $\rho(L)$ curve is monotonic at 7
of the 8 points. The one exception is the weakest-effect point tested (BA $z=4$,
$b=1.8$, original gain $0.12$): one of its 30 realizations is internally noisy at every
$L$ (replication-level std $0.23$--$0.35$, vs.\ $0.02$--$0.11$ elsewhere), consistent
with a genuinely near-bistable network rather than an artifact of too few replications,
and its gain comes out negative ($-0.06$), though the population-averaged curve at this
point is essentially flat between $L=3$ and $L=4$ rather than truly reversing. Spread
across realizations is modest everywhere (std $0.00$--$0.09$); a small fraction of
individual realizations (32 of 240, 13\%) show a local dip somewhere along
$L=1\to2\to3\to4$ without affecting the overall trend (Supplemental Material,
Sec.~S15).

Then, we ran an alternative study of between-network variability. For each (topology,
degree, inter-layer correlation, $L$) combination we selected representative
cooperative, defective, transition, and bistable $(b,\theta)$ points --- 111
combinations in total --- and reran each on 10 independently generated network
realizations. Between-network variability in $\langle\rho\rangle$ is consistently
smaller than the within-network variability already reported above, even in the
transition and bistable regimes where within-network variability is itself largest.
The fixed realization used throughout the paper is also representative of these values,
not merely low-variance relative to them: it correlates with the 10-realization average
at $r=0.99$ across all 111 points, and differs from it by more than $0.1$ at only 7
points, all in the bistable or transition regime, consistent with genuine bistability
--- a particular realization more or less likely to favor one attractor --- rather than a
biased choice of realization (Supplemental Material, Sec.~S15).

We also checked whether the cooperation gain from extending $L$ holds up under different
population sizes, Fermi noise, and update schemes, using the same 8 points as the first
study above: population sizes $N\in\{250,500,2000\}$, Fermi noise $K\in\{0.05,0.2\}$, and
an asynchronous Fermi update. The gain never changes sign under any condition tested, at
any of the 8 points. Sensitivity to $K$ is small throughout (mean $|\Delta|$
$0.02$--$0.03$). Sensitivity to $N$ falls off monotonically as $N$ grows past $1000$
(mean $|\Delta|$ $0.16$ at $N=250$, down to $0.02$ at $N=2000$), negative at 5 of the 8
points, consistent with finite-size noise weakening the effect rather than $N=1000$
having been tuned to an unusually favorable point. The asynchronous control shows small
deviations (mean $|\Delta|$ $0.04$), negative at 6 of the 8 points, with no systematic
pattern by topology (Supplemental Material, Sec.~S16).

\subsection{\label{sec:realnetwork}Robustness on an empirical social network}

The four network topologies studied so far are synthetic, raising the question of whether
long-range vigilance promotes cooperation on the structure of an actual social network.
We address this using the physician network collected by Coleman, Katz, and
Menzel~\cite{ColemanKatzMenzel1957,ColemanKatzMenzel1966} in their classic study of the
diffusion of a new drug among doctors in four Illinois towns---Peoria, Bloomington, Quincy,
and Galesburg---recoded by Burt~\cite{Burt1987} and distributed via the
\texttt{spatialprobit} R package~\cite{spatialprobit2024}.
This choice closes a methodological loop: Miranda \textit{et al.}~\cite{Miranda2024}
invoked physician networks to motivate the existence of indirect social influence, and
their measured decay coefficient is precisely what we used to calibrate $\lambda$
(Sec.~\ref{sec:model}); here we test the model on physician networks directly.
Three sociometric questions were asked of each physician: friendship (friends seen most
often socially), case discussion (physicians with whom cases or therapy are most often
discussed), and advice (where one usually turns for information or advice about therapy).
We assign friendship to $G_\mathrm{game}$, the natural substrate for everyday social
interaction, and advice to $G_\mathrm{vig}$: advice-seeking implies deference to a trusted
authority, which is closer to the vigilance-monitoring relation than peer case discussion.
The discussion network is not used.
Unlike the synthetic case, these are two genuinely different real relations rather than
independent realizations of the same random-graph model.

The four towns share no edges with one another in either relation, so we treat them as four
independent networks, exactly as the synthetic topologies above.
Within each town we take the giant component of the (symmetrized) friendship network as the
node set and restrict the advice network to those same nodes; physicians with zero advice-degree
receive $I_i=0$, as in the uncorrelated synthetic multiplex.
The resulting networks are small and sparse compared to the synthetic case
($N=34$--$110$, mean degree $3.2$--$4.2$, comparable to our sparse $z=4$ topologies rather
than $z=16$), a real-world contrast we report explicitly rather than treat as a limitation.
We use $\lambda=0.65$ here---rather than the illustrative $\lambda=0.5$ used above---taken
directly from the empirical decay coefficient $c_2\approx0.65$ measured by
Ref.~\cite{Miranda2024} for the second circle of influence, the more precisely estimated of
their two reported indirect-influence coefficients.

\begin{figure}[t]
  \centering
  \includegraphics[width=\columnwidth]{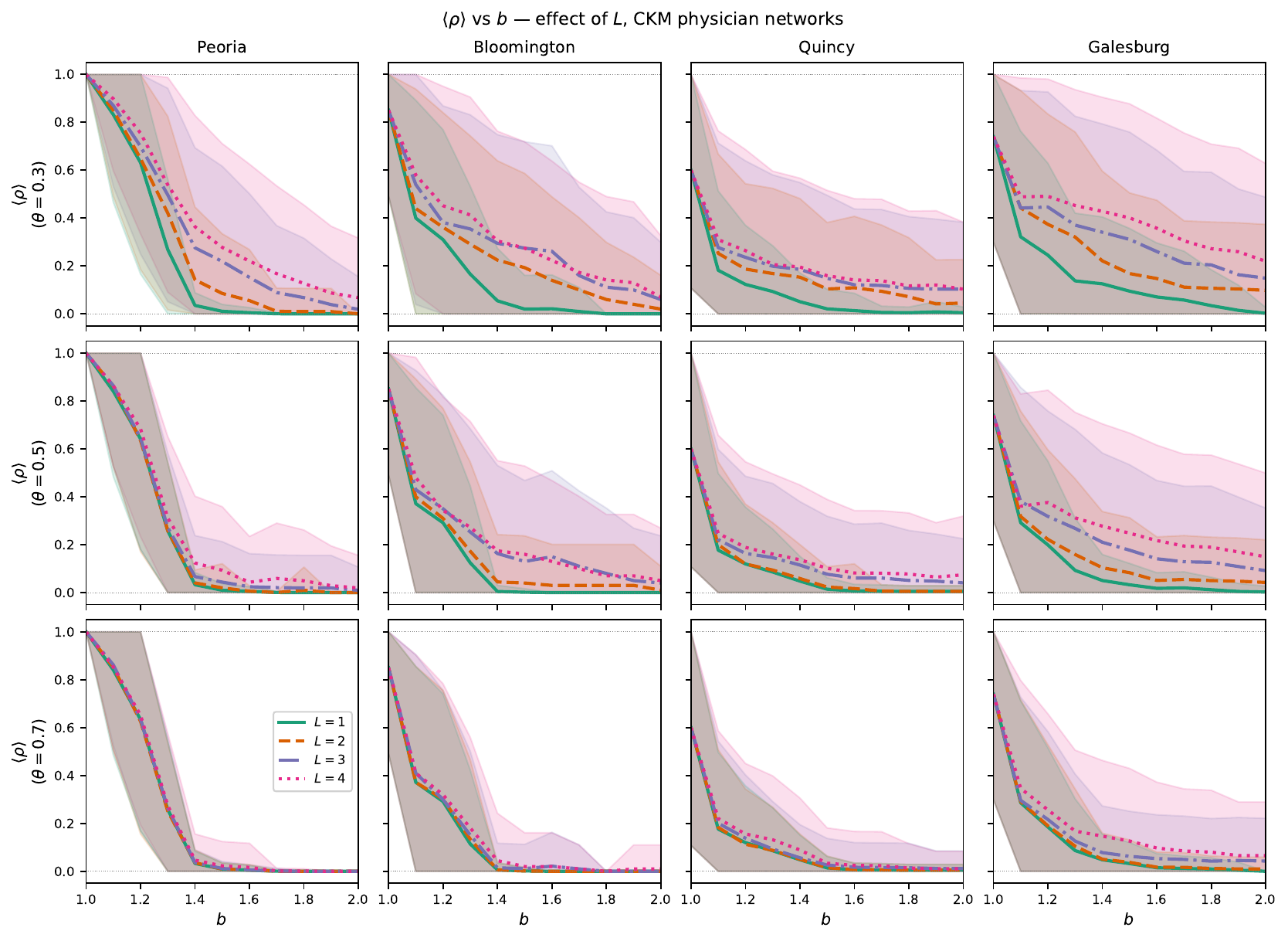}
  \caption{Stationary cooperation fraction $\langle\rho\rangle$ versus temptation $b$ at
    fixed $\theta=0.3$ (top), $0.5$ (middle), and $0.7$ (bottom) for the four CKM physician
    networks (Fermi rule, $\lambda=0.65$).
    Columns correspond to the four towns: Peoria ($N=110$), Bloomington ($N=46$), Quincy
    ($N=38$), Galesburg ($N=34$).
    Curves: $L=1$ (teal, solid), $L=2$ (orange, dashed), $L=3$ (purple, dash-dot), $L=4$
    (magenta, dotted).
    Shaded bands indicate $\pm1$ standard deviation across 100 replications.
    Full $(b,\theta)$ heatmaps in Fig.~S11.}
  \label{fig:realnetwork}
\end{figure}

Figure~\ref{fig:realnetwork} shows the stationary cooperation fraction versus temptation
$b$ at three threshold values for all four towns; full $(b,\theta)$ heatmaps are in the
Supplemental Material (Fig.~S11).
Extending the vigilance range promotes cooperation in every town, reproducing the
qualitative result found on synthetic networks despite the very different size, sparsity,
and source of these graphs.
At $\theta=0.3$, the critical temptation at which cooperation collapses to
$\langle\rho\rangle<0.05$ starts at $b=1.4$--$1.7$ ($L=1$) and exceeds the simulated range
($b>2.0$) by $L=4$ in all four towns, though at different rates: in Galesburg the
collapse boundary already exits the simulated range at $L=2$; in Bloomington and Quincy
at $L=3$; in Peoria, the largest town, only at $L=4$, with $L=1\to3$ shifting the boundary
from $b=1.4$ to $b=1.9$.
The size of the gain varies with town: at $\theta=0.3$, $b=1.5$, extending from $L=1$ to
$L=4$ raises $\langle\rho\rangle$ from $0.020$ to $0.275$ in Bloomington and from $0.095$ to
$0.401$ in Galesburg---the smallest town but the one with the highest mean friendship
degree---while the gain is comparatively modest in Quincy ($0.021$ to $0.159$).

As in the synthetic networks, the benefit of extending $L$ is concentrated at
low-to-intermediate vigilance threshold and nearly disappears at high threshold: at
$\theta=0.7$, the mean cooperation gain from $L=1$ to $L=4$, averaged over $b$, is below
$0.07$ in every town (compared with gains of $0.11$--$0.23$ at $\theta=0.3$ for the same
comparison), because few physicians reach a demanding influence threshold regardless of how
many circles are summed.
Variance across replications is markedly higher than in the synthetic networks of
comparable sparsity (mean $\sigma\approx0.19$--$0.30$ here versus $\lesssim0.1$ in most of
the ER $z=4$ parameter space), consistent with the stronger role of demographic
stochasticity in networks an order of magnitude smaller than $N=1000$; even the neutral
case $b=1.0$, where cooperators and defectors are payoff-equivalent, settles below full
cooperation in three of the four towns ($\langle\rho\rangle=0.60$--$0.85$) purely from stochastic fluctuations in the update dynamics, which are stronger at
smaller $N$, whereas synthetic networks at $N=1000$ remain at
$\langle\rho\rangle\gtrsim0.99$ under the same condition.
This added noise does not, however, eliminate the systematic effect of $L$: the $L=1\to4$
gain exceeds its own propagated standard error at roughly half to three-quarters of the
points in each town's collapse region ($47$--$70\%$, lowest in Peoria and highest in
Galesburg), rather than being attributable to noise throughout.

\section{\label{sec:conclusions}Conclusions}

Beyond coupling cooperation to vigilance, already established in~\cite{Pereda2016}, this
model's conceptual contribution is to introduce distance-resolved, multi-hop social pressure
into an evolutionary-game framework, bridging the literature on structured and multilayer
cooperation with studies of multi-step social influence and complex contagion.

We extended the monitoring model of~\cite{Pereda2016} to allow vigilance to propagate
beyond direct neighbors, reaching $L$ circles of influence with a geometrically decaying
weight $\lambda^{d-1}$.
The Fermi update rule qualitatively reproduces the main structure of the replicator phase
diagram of~\cite{Pereda2016} at $L=1$ (Sec.~\ref{sec:REPvsFermi}), confirming that our
extension preserves the original model's qualitative behavior in its original limit, even
though the two rules differ quantitatively in topology-dependent ways.
Our central result is that extending vigilance beyond direct neighbors substantially
promotes cooperation, though how quickly the gain accrues is itself topology dependent: in
sparse Erd\H{o}s--R\'enyi networks the $L=1\to2$ step alone already recovers most of it,
while sparse Barab\'asi--Albert networks need a third circle before the gain is similarly
complete; the effect is strongest precisely where the original model is weakest---sparse
networks, where direct-neighbor vigilance alone struggles to sustain cooperation at high
temptation.
This is consistent with the empirical observation that social influence on behavior persists
to at least three degrees of separation~\cite{Christakis2007,Fowler2008} and with the
empirical decay coefficients of indirect influence reported by~\cite{Miranda2024}: cooperation
does not require an implausibly wide monitoring radius, only that the signal from the
second and third circles---whose decay coefficients are empirically consistent with the
geometric kernel used here---be allowed to reach the individual at all.

This reach matters because vigilance, as modeled here through the Watts threshold
rule~\cite{Watts2002}, is a complex contagion~\cite{Centola2007}: a single vigilant
neighbor is rarely enough to push an agent's influence index past $\theta$, but
reinforcement accumulated from a wider neighborhood can. Extending $L$ is therefore not
merely a quantitative enlargement of the monitoring radius but a qualitative change in
how easily the threshold is reached, which is why the cooperation gain we report is
concentrated at low-to-intermediate $\theta$ and vanishes once the threshold becomes too
demanding for any finite circle to satisfy, in both the synthetic and the real
networks studied here.
Whether this reinforcement is functionally useful, however, depends on who is
doing the reinforcing: cooperation only benefits from a wider vigilance radius when the
vigilance and game layers are aligned, so that the agents whose watchfulness lowers an
individual's temptation are the same agents that individual actually plays against.
When the two layers are uncorrelated, this alignment breaks and the benefit of extending
$L$ is largely lost---except in Barab\'asi--Albert networks, where the shared hub
structure of any two realizations of the same degree distribution partially preserves the
correlation that Erd\H{o}s--R\'enyi networks lack entirely.
This distinction maps naturally onto contemporary social life, where the people one
monitors online are not always the people one cooperates with offline; our results suggest
that this misalignment, not merely the existence of a wide monitoring network, can blunt
its cooperative benefit.

Testing the model's robustness on the CKM physician network~\cite{ColemanKatzMenzel1957,ColemanKatzMenzel1966}
showed that this mechanism is not an artifact of idealized random-graph topologies: the
same qualitative pattern---cooperation gains concentrated at low threshold, vanishing at
high threshold---reproduces on a real, sparse, historically documented social network at
an order of magnitude smaller $N$ than our synthetic networks, despite the
correspondingly larger role of demographic noise.
The sensitivity analysis further revealed that the decay parameter $\lambda$ does not act
uniformly across topologies: sparse networks respond smoothly to $\lambda$, while dense,
hub-dominated networks can switch sharply between a low- and a high-cooperation regime as
$\lambda$ crosses a threshold between $0.5$ and $0.75$.

Our model retains the original's simplifying assumptions: vigilance and strategy are
binary rather than graded, and monitoring carries no direct cost to the monitor.
Relaxing either assumption is a natural next step, particularly because a cost of
vigilance would couple the threshold dynamics studied here to the second-order free-rider
problem inherent in any costly monitoring scheme.

Our extension to $L$ circles of influence bears a conceptual connection to the emerging
literature on higher-order interactions in complex systems~\cite{Battiston2020,Battiston2021,Battiston2025}:
both frameworks acknowledge that pairwise network connections are insufficient to capture
the full complexity of social influence, which often involves groups rather than dyads.
As recently argued, higher-order structures reveal mechanisms of social contagion and
cooperation that are invisible in dyadic models~\cite{Battiston2025}.
Exploring vigilance cascades on hypergraphs or simplicial complexes---where group-level
monitoring is an explicit structural feature---represents a natural avenue for future work.

While our model focuses on the cooperative benefits of long-range vigilance, real
monitoring mechanisms carry individual and social costs---including privacy and
autonomy---not captured in the present framework, and whose study is left for future work.

\bibliography{refs}

\end{document}